\documentclass[aps,twocolumn]{revtex4}

\begin{document}

\title
{Statistical Mechanics of Quantum-Classical Systems with Holonomic Constraints}

\author{Alessandro Sergi
\footnote{E-mail: asergi@unime.it}}
\affiliation{
Dipartimento di Fisica, 
Universit\'a degli Studi di Messina,
Contrada Papardo 98166 Messina, Italy
}

\begin{abstract}
The statistical mechanics of quantum-classical systems
 with holonomic constraints
is formulated rigorously by unifying the classical Dirac bracket and the
quantum-classical bracket in matrix form.
 The resulting Dirac quantum-classical
theory, which conserves the holonomic constraints exactly, is then used
to formulate time evolution and statistical mechanics.
The correct momentum-jump approximation for constrained system arises
naturally from this formalism.
Finally, in analogy with what was found in the classical case, it
is shown that the rigorous linear
response function of constrained quantum-classical systems
contains non-trivial additional  terms which are absent
in the response of unconstrained systems.
\end{abstract}

\maketitle


\section{Introduction}

Quantum-classical formalisms~\cite{qc-bracket}
 are computationally useful approximations of full quantum mechanics.
In the past few years, a mathematical formalism, which permits to derive
surface-hopping schemes from full quantum mechanics by means of a number of
controlled approximations, has been proposed~\cite{kakkacicco}.
Most notably a consistent statistical theory of quantum-classical systems
has been first introduced~\cite{nciccokakka}
and then used in order to formulate the theory of 
nonadiabatic rate constants~\cite{sergi}.
Numerical calculations of quantum-classical rate
constants in nonadiabatic chemical reactions have been performed for some 
model systems which are relevant for condensed phase~\cite{sergi,sergi2}.
Despite some numerical problems
with long-time integration of surface-hopping trajectories, this approach
holds promise for the study of more realistic systems with few quantum degrees of 
freedom but many classical particles. As a matter of fact, just recently, 
Hanna and Kapral~\cite{hannakakka} reported the results
 of a quantum-classical calculation of the
nonadiabatic rate constant for the proton transfer process occurring in a two-atom
complex immersed in a classical bath of diatomic molecules, which were 
represented by means of cartesian coordinates and holonomic constraints.
This example shows how, within the quantum-classical formalism 
of Ref.~\cite{kakkacicco},
one can easily model classical baths which are, in principle, as complex 
as the state of art atomistic representation of a protein by means of 
holonomic constraints and force fields (for an example see Refs.~\cite{serma}).
However, in order to do so rigorously, one must generalize the quantum-classical
formalism of Ref.~\cite{kakkacicco}
so that it can describe classical baths with holonomic
constraints. This issue was not addressed in Ref.~\cite{hannakakka}.
Thus, the motivation
of the present paper is to formulate a consistent statistical mechanics
of quantum-classical systems when holonomic constraints are employed 
to model the classical bath.
Linear response theory for quantum-classical constrained systems
is also given because it is particularly relevant for the calculation
of nonadiabatic rate constants.

A rigorous formulation of linear response theory for fully classical systems
with holonomic constraints has been introduced only recently~\cite{bdispettosa}
by means of a re-interpretation of the formalism originally developed
by Dirac~\cite{dirac}. This formalism~\cite{bdispettosa}, based on a
generalization~\cite{b1,b2} of the symplectic structure of classical
brackets in phase space~\cite{goldstein,mccauley,morrison}, showed that
unusual terms may appear in the response function of a classical system
with holonomic constraints. In light of this result, the extension of the
theory of Refs.~\cite{bdispettosa} to the quantum-classical case is a subtle issue.
The identification of the symplectic structure of quantum commutators~\cite{b3}
and the generalization of this structure to introduce a non-Hamiltonian
quantum bracket~\cite{b3}, of which the quantum-classical bracket of
Refs.~\cite{qc-bracket,kakkacicco} is a particular realization, has also
been proposed just recently and used to introduce quantum-classical
Nos\'e-Hover dynamics~\cite{b3}.

In the present work, the non-Hamiltonian commutator of Ref.~\cite{b3} and the 
classical Dirac bracket of Refs.~\cite{bdispettosa} are combined in order
to formulate the statistical theory of quantum-classical systems
with holonomic constraints. A Dirac quantum-classical bracket,
which exactly conserves constraints, is easily introduced
and used consistently to define the dynamics and the statistical
mechanics of quantum-classical systems with holonomic constraints. 
In particular, it will be shown how the momentum-jump approximation 
must be modified in order to consider correctly the back-reaction
of the quantum degrees of freedom on the constrained classical
momenta. The stationary constrained quantum-classical density matrix
will be derived and linear response theory formulated.
As already observed in the classical case~\cite{bdispettosa},
non-trivial terms, associated to the action of the perturbation
on the phase space measure of the unperturbed constrained system
and the phase space compressibility introduced by
the perturbation itself, must be considered in general.
It results from the derivation that these additional terms
are zero if quantum-classical variables, to which the perturbation
is coupled, depend only from particle positions. In this case, the
rate formulas derived in Refs~\cite{sergi,sergi2,hannakakka}
can be applied to a constrained systems with the only
additional requirement of using the correct constrained stationary density matrix.
However, different perturbations could require to evaluate
all the terms in the response function of quantum-classical constrained systems
and one must be aware of their existence.

If the work presented in this paper is considered together with that
of Ref.~\cite{b3}, where non-Hamiltonian quantum commutators
where introduced and used to introduce quantum-classical Nos\'e
dynamics, it is readily realized that a unified formalism
for defining generalized dynamics and statistical mechanics
in quantum-classical systems is now available.
There are reasonable expectations that this formalism
could be used in the future in order to attack the
problem of long-time numerical integration of quantum-classical dynamics.

This paper is organized as follows. In Sec~\ref{sec:dirac} Dirac formalism,
as given in Ref.~\cite{bdispettosa}, is shortly summarized. In Sec.~\ref{sec:gqm},
the results of Ref.~\cite{b3} are quickly presented by showing the 
generalized symplectic structure of quantum-classical brackets,
which are a particular realization of non-Hamiltonian quantum commutators. 
In Sec.~\ref{sec:qcd-constr}
the matrix structure of quantum-classical bracket,
is used to combine it with Dirac bracket.
In such a way the Dirac quantum-classical formalism is introduced
and equations of motion which preserves the constraints exactly are given.
In Sec.~\ref{sec:ad-basis} such equations of motion are represented
in the adiabatic basis and the correct momentum-jump approximation
for a constrained system is derived easily.
In Sec.~\ref{sec:stationary} the quantum-classical stationary density matrix
for a system with holonomic constraints is derived from the
Dirac quantum-classical bracket.
In Sec.~\ref{sec:lrt} rigorous linear response for constrained systems is introduced.
Finally, conclusions are given in Sec.~\ref{sec:concl}.


\section{Dirac Bracket for classical systems 
with holonomic constraints}\label{sec:dirac}

Consider the phase space point $X=(R,P)$, where $R$ and $P$
are coordinates and momenta, respectively, of the system under 
consideration.
Let
\begin{equation}
\mbox{\boldmath$\cal B$}^s=
\left[\begin{array}{cc}{\bf 0} & {\bf 1}\\ -{\bf 1} &{\bf 0}\end{array}\right]
\label{B}
\end{equation}
be the symplectic matrix, then it is well known~\cite{goldstein,mccauley}
that Poisson brackets can be written as
\begin{equation}
\{a,b\}=\sum_{i,j=1}^{2N}\frac{\partial a}{\partial X_i}{\cal B}_{i j}^s
\frac{\partial b}{\partial X_j}\;,
\label{eq:symplectic-bracket}
\end{equation}
where $a(X)$ and $b(X)$  are two arbitrary phase space functions
and $2N$ is the dimension of phase space.
The structure of Eq.~(\ref{eq:symplectic-bracket}) has been
exploited to introduce Hamiltonian
non-canonical flows~\cite{morrison}, non-Hamiltonian~\cite{b1,b2} and 
constrained flows~\cite{bdispettosa}.
Here it is summarized how to generalize the structure of 
Eq.~(\ref{eq:symplectic-bracket}) in order to define equations of motion
for systems with holonomic constraints.

Consider a system with Hamiltonian ${\cal H}_0$ and a set of 
phase space constraints
\begin{equation}
\xi_{\alpha}(X)=0\quad \alpha=1,\dots, 2 l~.
\end{equation}
Following Dirac~\cite{dirac}, one can introduce the matrix
\begin{equation}
C_{\alpha\beta}=\{\xi_{\alpha},\xi_{\beta}\}=
\sum_{i,j=1}^{2N}\frac{\partial\xi_{\alpha}}{\partial X_i}
{\cal B}_{i j}^c
\frac{\partial\xi_{\beta}}{\partial X_j}
\label{matrixC}
\end{equation}
and its inverse $\left({\bf C}^{-1}\right)_{\alpha\beta}$,
where $\alpha,\beta=1,...,2l$.
Note that, when considering Poisson brackets of the constraints,
the convention of first evaluting the brackets and then imposing the
constraints must be followed~\cite{dirac}.
By defining an antisymmetric matrix $\mbox{\boldmath$\cal B$}^D$
\begin{equation}
{\cal B}_{i j}^D(X)={\cal B}_{i j}^s-\sum_{k,m=1}^{2N}
\sum_{\alpha,\beta=1}^{2l}
{\cal B}_{i k}^s\frac{\partial\xi_{\alpha}}{\partial X_k}
\left({\bf C}^{-1}\right)_{\alpha\beta}
\frac{\partial\xi_{\beta}}{\partial X_m}{\cal B}_{m j}^s
\;, \label{B^D}
\end{equation}
the Dirac bracket can be introduced as
\begin{equation}
\{a,b\}_D= \sum_{i,j=1}^{2N}\frac{\partial a}{\partial X_i}
{\cal B}_{i j}^D \frac{\partial b}{\partial X_j}\;.  \label{dirac-bracket}
\end{equation}
Equation~(\ref{dirac-bracket})  
was originally given by Dirac in the equivalent form~\cite{dirac}
\begin{equation}
\{a,b\}_D=\{a,b\}-\sum_{\alpha,\beta=1}^{2l}\{a,\xi_{\alpha}\}
\left({\bf C}^{-1}\right)_{\alpha\beta}
\{\chi_{\beta},b\}\;.
\label{eq:diracequiv}
\end{equation}
Constrained phase space flows are then defined by 
\begin{equation}
\dot{a}=\{a,{\cal H}_0\}_D\;,\label{dirac_flow}
\end{equation}
so that the Hamiltonian and any function
of the constraints is conserved
because, considering an arbitrary function $f(\xi_{\alpha})$ of the constraints,
it is easy to verify that $\{f(\xi_{\sigma}),{\cal H}_0\}_D=0$.
The above formalism due to Dirac has been specialized
to non-relativistic systems with holonomic constraints~\cite{bdispettosa}.

In order to see how this can be achieved,
consider a system with a number $l$ of holonomic constraints
in configuration space $\sigma_{\alpha}(\{R\})=0$, $\alpha=1,...,l$.
Consider also the following additional constraints 
$\dot{\sigma}_{\alpha}(\{R,\dot{R}\})=
\sum_{i=1}^N(\partial\sigma_{\alpha}/\partial R)\cdot
P_i/M=0$,~ $\alpha=1,...,l$, where 
$M$ are the particle masses.
The whole set of constraints can be denoted as
\begin{equation}
(\xi_1,...,\xi_l,\xi_{l+1},...,\xi_{2l})=
(\sigma_1,...,\sigma_l,\dot{\sigma}_1,...,\dot{\sigma}_l)\;.
\label{eq:wholeconstr}
\end{equation}
Defining the matrices
\begin{eqnarray}
\Gamma_{\alpha\beta} &=&
\sum_{i,k=1}^{N}
\left( \frac{1}{M}\frac{\partial\sigma_{\beta}}{\partial R}
\cdot \frac{\partial^2\sigma_{\alpha}}{\partial R\partial R}
\cdot\frac{P}{M} \right.\nonumber \\
&&\left.
- \frac{1}{M} \frac{\partial\sigma_{\alpha}}{\partial R}\cdot
\frac{\partial^2\sigma_{\beta}}{\partial R\partial R} \cdot\frac{P}{M} 
\right)
 \;,
\end{eqnarray}
and
\begin{equation}
Z_{\alpha\beta}=\sum_i\frac{1}{M}
\frac{\partial\sigma_{\alpha}}{\partial R}\cdot
\frac{\partial\sigma_{\beta}}{\partial R} \;.
\end{equation}
Then one finds
\begin{equation}
{\bf C}= \left[
\begin{array}{cc} 0 & {\mathbf Z} \\
-{\mathbf Z} & \mbox{\boldmath$\Gamma$}
\end{array} \right] \;,
\end{equation}
and
\begin{equation}
{\bf C}^{-1}= \left[
\begin{array}{cc} {\bf Z}^{-1}\mbox{\boldmath$\Gamma$}{\bf Z}^{-1} &
-{\bf Z}^{-1} \\ {\bf Z}^{-1} & 0
\end{array} \right] \;.
\label{C^-1}
\end{equation}
The time evolution of the phase space point
under the constrained dynamics is then given by the Dirac bracket.
The explicit equations of motion are 
\begin{eqnarray}
\dot{R}&=&\{R,{\cal H}_0\}_D = \frac{P}{M}
\;, \label{r-flow}
\\
\dot{P}&=&\{P,{\cal H}_0\}_D
 = F-\sum_{\alpha=1}^{l}
\lambda_{\alpha}
\frac{\partial\sigma_{\alpha}}{\partial R}
\;,
\label{p-flow}
\end{eqnarray}
where $F$ are the forces acting on the particles.
The $\lambda_{\alpha}$ are
the exact Lagrange multipliers 
\begin{eqnarray}
\lambda_{\alpha}&=&\sum_{\beta=1}^{l}Z_{\alpha\beta}^{-1}
\{\dot{\sigma}_{\beta},{\cal H}_0\}
\;,\nonumber\\
&=&\sum_{\beta=1}^l\left(
\frac{P}{M}\otimes\frac{P}{M}\cdot
\frac{\partial^2\sigma_{\beta}}{\partial R\partial R}
+\frac{F}{M}\cdot\frac{\partial\sigma_{\beta}}{\partial R}\right)
\;.
\label{lambda_alpha}
\end{eqnarray}

\section{GENERALIZED QUANTUM MECHANICS
and QUANTUM-CLASSICAL BRACKETS}\label{sec:gqm}

Consider now the realm of quantum mechanics
and an arbitrary set of quantum variables
$\hat{\chi}_A$, $A=1,...,n$.
The commutator
$[\hat{\chi}_A,\hat{\chi}_N]_-=\hat{\chi}_A\hat{\chi}_N
-\hat{\chi}_N\hat{\chi}_A$ ($A,N=1,...,n$)
was written in symplectic form~\cite{b3}
using the matrix defined in Eq.~(\ref{B}):
\begin{equation}
[\hat{\chi}_A,\hat{\chi}_N]=
\left[\begin{array}{cc}\hat{\chi}_A & \hat{\chi}_N\end{array}\right]
\cdot \mbox{\boldmath$\cal B$}^s \cdot
\left[\begin{array}{c}\hat{\chi}_A \\ \hat{\chi}_N\end{array}\right].
\label{eq:quantum-algebra}
\end{equation}
Considering the Hamiltonian operator of the system $\hat{H}_0$,
the laws of motion in the Heisenberg picture are written as
\begin{equation}
 \frac{d\hat{\chi}_A}{dt}=\frac{i}{\hbar}
\left[\begin{array}{cc} \hat{H}_0 & \hat{\chi}_A \end{array}\right]
\cdot \mbox{\boldmath$\cal B$}^s\cdot
\left[\begin{array}{c} \hat{H}_0 \\ \hat{\chi}_A\end{array}\right]
=i\hat{\cal L}\hat{\chi}_A,
\label{qlm}
\end{equation}
where the Liouville operator
\begin{equation}
i \hat{\cal L}=\frac{i}{\hbar}
\left[\begin{array}{cc} \hat{H}_0 & \ldots \end{array}\right] \cdot
\mbox{\boldmath$\cal B$}^s\cdot
\left[\begin{array}{c} \hat{H}_0 \\ \ldots \end{array}\right]
\end{equation}
has also been introduced.

As it was showed in Ref.~\cite{b3}, the laws of quantum
mechanics can be generalized by defining the antisymmetric matrix operator
\begin{equation}
\mbox{\boldmath$\cal D$}=
\left[\begin{array}{cc}0 & \hat{\zeta}\\ -\hat{\zeta} & 
0\end{array}\right]\;,
\label{D}
\end{equation}
with $\hat{\zeta}$ arbitrary operator or c-number,
and defining non-Hamiltonian quantum brackets (commutators) as
\begin{equation}
[\hat{\chi}_A,\hat{\chi}_N]_-=
\left[\begin{array}{cc}\hat{\chi}_A & \hat{\chi}_N\end{array}\right]
\cdot\mbox{\boldmath$\cal D$}\cdot
\left[\begin{array}{c}\hat{\chi}_A \\ \hat{\chi}_N\end{array}\right]\;.
\label{eq:gen-quantum-algebra}
\end{equation}
Generalized equations of motion are then defined by
\begin{equation}
 \frac{d\hat{\chi}_A}{dt}=\frac{i}{\hbar}
\left[\begin{array}{cc} \hat{H}_0 & \hat{\chi}_A \end{array}\right]
\cdot\mbox{\boldmath$\cal D$}\cdot
\left[\begin{array}{c} \hat{H}_0 \\ \hat{\chi}_A\end{array}\right]
=i\hat{\cal L}\hat{\chi}_A.
\label{eq:gen-qlm}
\end{equation}

Quantum-classical systems are defined in terms of
quantum variables $\hat{\chi}_A$ in a Hilbert space
which depends from the classical phase space
point $X$, i.e. $\hat{\chi}_A(X)$.
Time evolution in the Hilbert space and in classical phase
space are coupled consistently.
The energy $E_0$ of such systems is defined in terms of
a quantum-classical Hamiltonian operator $\hat{H}_0=\hat{H}_0(X)$ 
so that $E_0={\rm Tr}'\int dX \hat{H}_0(X)$.
The dynamical evolution of a quantum-classical operator $\hat{\chi}(X)$
is given by the quantum-classical bracket
which is defined
in terms of the commutator and the symmetrized
sum of Poisson bracket~\cite{qc-bracket,kakkacicco}.
Following Refs.~\cite{b1,b2}, in Ref.~\cite{b3}
the quantum-classical bracket was easily casted in matrix
form as a non-Hamiltonian commutator.
To this end, one can consider the operator $\Lambda$
defined in such a way that applying its negative on any pair
of quantum-classical operators $\hat{\chi}_A(X)$ and
 $\hat{\chi}_N(X)$ their Poisson bracket is obtained
\begin{equation}
\{\hat{\chi}_A,\hat{\chi}_N\}=
-\hat{\chi}_A(X)\hat{\Lambda}\hat{\chi}_N(X)=
\sum_{i,j=1}^{2N}\frac{\partial \hat{\chi}_A}{\partial X_i}{\cal B}_{i j}^s
\frac{\partial \hat{\chi}_N}{\partial X_j}.
\label{Lambda}
\end{equation}
If one defines
\begin{equation}
\zeta^{qc}=1+\frac{\hbar\Lambda}{2i}
\end{equation}
in Eq.~(\ref{D}) then a new matrix operator
is introduced as
\begin{equation}
\mbox{\boldmath$\cal D$}^{qc}=\zeta^{qc}\mbox{\boldmath$\cal B$}^s
\end{equation}
and the quantum-classical law of motion can be written as
\begin{equation}
\partial_t \hat{\chi}_A=\frac{i}{\hbar}
\left[\begin{array}{cc} \hat{H}_0 & \hat{\chi}_A \end{array}\right]
\cdot\mbox{\boldmath$\cal D$}^{qc}\cdot
\left[\begin{array}{c} \hat{H}_0 \\ \hat{\chi}_A \end{array}\right]
=(\hat{H}_0,\hat{\chi}_A)
=i\hat{\cal L}\hat{\chi}_A,
\label{qclm}
\end{equation}
where the last equality introduces the quantum-classical Liouville operator
in terms of the quantum-classical bracket.

\section{QUANTUM-CLASSICAL DYNAMICS
WITH HOLONOMIC CONSTRAINTS} \label{sec:qcd-constr}

Consider a quantum-classical system with Hamiltonian $\hat{H}_0(X)$
and a set of holonomic constraints, as specified by Eq.~(\ref{eq:wholeconstr}),
acting on the classical bath variables $X$.
Then introduce an operator $\Lambda^D$ such that it can be used to give the negative
of the Dirac bracket of two arbitrary quantum-classical variables
\begin{equation}
\{\hat{\chi}_A,\hat{\chi}_N\}_D=
-\hat{\chi}_A(X)\hat{\Lambda}^D\hat{\chi}_N(X)=
\sum_{i,j=1}^{2N}
\frac{\partial \hat{\chi}_A}{\partial X_i}{\cal B}_{i j}^D
\frac{\partial \hat{\chi}_N}{\partial X_j}\;,
\label{LambdaD}
\end{equation}
where $\mbox{\boldmath $\cal B$}^D$ has been defined in Eq.~(\ref{B^D}).
Using $\Lambda^D$, one can define
\begin{equation}
\zeta^D=1+\frac{\hbar\Lambda^D}{2i}\;,
\label{zeta^D}
\end{equation}
and the constrained matrix operator
\begin{equation}
\mbox{\boldmath$\cal D$}^D=\zeta^D\mbox{\boldmath$\cal B$}^s \;.
\label{D^D}
\end{equation}
Using $\zeta^D$ and $\mbox{\boldmath$\cal D$}^D$, defined in Eq.~(\ref{zeta^D})
and~(\ref{D^D}) respectively, equations of motion for quantum-classical 
systems with holonomic constraints are defined by
\begin{equation}
\partial_t \hat{\chi}_A=\frac{i}{\hbar}
\left[\begin{array}{cc} \hat{H}_0 & \hat{\chi}_A \end{array}\right]
\cdot\mbox{\boldmath$\cal D$}^D\cdot
\left[\begin{array}{c} \hat{H}_0 \\ \hat{\chi}_A \end{array}\right]
=(\hat{H}_0,\hat{\chi}_A)_D
=i\hat{\cal L}^D\hat{\chi}_A\;.
\label{qclm-dirac}
\end{equation}
In Equation~(\ref{qclm-dirac}), the Dirac quantum-classical bracket
$\{\ldots ,\ldots\}_D$
and the Dirac Liouville operator $i{\cal L}^D$ have been introduced.
Equation~(\ref{qclm-dirac}) is one of the main results of this work.
It introduces the correct algebraic quantum-classical evolution
for systems with holonomic constraints. In the following, all the other
results of this paper will be derived from this equation.

One can derive the fist consequence of Eq.~(\ref{qclm-dirac})
by considering the time evolution on an arbitrary function
$f(\xi_{\gamma})$ of the holonomic constraints.
This is of course given by the Dirac quantum-classical bracket
$(\hat{H}_0,f(\xi_{\gamma}))_D$ which can be written explicitly as
\begin{eqnarray}
\left(\hat{H}_0,f(\xi_{\gamma})\right)_D
&=&
\frac{i}{\hbar}\left[\begin{array}{cc} \hat{H}_0 & f(\xi_{\gamma})
\end{array}\right]\nonumber \\
&\cdot&
\left[\begin{array}{cc} 0 & 1+ \frac{\hbar\Lambda^D}{2i} \\
-1-\frac{\hbar\Lambda^D}{2i} & 0 \end{array}\right]\cdot
\left[\begin{array}{c}\hat{H}_0\\f(\xi_{\gamma})\end{array}\right]
\nonumber \\
&=&\frac{i}{\hbar}\left[\hat{H}_0,f(\xi_{\gamma})\right]
-\frac{1}{2}\left(\{\hat{H}_0,f(\xi_{\gamma})\}_D
\right. \nonumber\\
&&\left.
-\{f(\xi_{\gamma}),\hat{H}_0\}_D\right)\;.
\label{qcDconstr}
\end{eqnarray}
Equation~(\ref{qcDconstr}) shows explicitly that the Dirac quantum-classical
bracket is defined in terms of the symplectic commutator
(in the secon equality, the first term on the right hand side)
of two variables minus one half of the antisymmetric
combination of classical Dirac brackets.
The symplectic commutator of the quantum-classical Hamiltonian $\hat{H}_0$
with the arbitrary function $f(\xi_{\gamma})$ is identically zero
because, by hypothesis, the phase space constraints
 $\xi_{\gamma}$, $\gamma=1,\ldots,2l$,
do not involve quantum degrees of freedom. Then,
in order to see what is the effect of the quantum-classical Dirac bracket
on the arbitrary function of the constraints $f(\xi_{\gamma})$,
there remains to be considered 
the action of the classical Dirac bracket $\{\ldots,\ldots\}_D$.
For example, take
\begin{eqnarray}
\{\hat{H}_0,f(\xi_{\gamma})\}_D
&=&\{\hat{H}_0,f(\xi_{\gamma})\}\nonumber\\
&-&
\{\hat{H}_0,\xi_{\beta}\}({\bf C})^{-1}_{\beta\alpha}
\{\xi_{\alpha},f(\xi_{\gamma})\}
\nonumber \\
&=&\{\hat{H}_0,f(\xi_{\gamma})\}
\nonumber\\
&-&\sum_{\alpha,\beta,\mu=1}^{2l}
\{\hat{H}_0,\xi_{\beta}\}({\bf C})^{-1}_{\beta\alpha}
C_{\alpha\mu}\frac{\partial f(\xi_{\gamma})}{\partial \xi_{\mu}}
\nonumber\\
&=&\{\hat{H}_0,f(\xi_{\gamma})\}-\sum_{\mu=1}^{2l}
\{\hat{H}_0,\xi_{\mu}\}\frac{\partial f(\xi_{\gamma})}{\partial \xi_{\mu}}
=0\;. \nonumber\\
\end{eqnarray}
In the same manner, one finds that $\{f(\xi_{\gamma}),\hat{H}_0\}_D=0$
so that, as desired, the Dirac quantum-classical bracket
leaves invariant any function of the holonomic constraints.

\section{REPRESENTATION OF THE QUANTUM-CLASSICAL DIRAC BRACKET
IN THE ADIABATIC BASIS}\label{sec:ad-basis}

Equation~(\ref{qclm-dirac}) is an algebraic equation. In order to actually
perform numerical calculations one needs to introduce a basis.
The adiabatic basis is particularly suited to represent
quantum-classical equations of motion~\cite{kakkacicco},
discuss quantum-classical statistical mechanics~\cite{nciccokakka},
obtain surface-hopping algorithms~\cite{sergi,sergi2,hannakakka},
and generalize quantum-classical dynamics in order to have
constant temperature dynamics on the classical bath degrees of freedom~\cite{b3}.
To define this basis, consider the following specific form of the
quantum-classical Hamiltonian 
\begin{eqnarray}
\hat{H}_0&=&\frac{P^2}{2M}+\hat{K}+\hat{\Phi}(R)
\nonumber\\
&=&\frac{P^2}{2M}+\hat{h}(R)\;,
\end{eqnarray}
where $\hat{K}$ is the kinetic energy operator 
of the quantum degrees of freedom and $\hat{\Phi}(R)$
is the potential energy operator coupling quantum and classical
variables. It is well known that adiabatic states are defined
by the eigenvalue equation
\begin{equation}
\hat{h}(R)|\alpha;R\rangle=E_{\alpha}(R)|\alpha;R\rangle \;.
\end{equation}
Before finding the representation in the adiabatic basis,
Eq.~(\ref{qclm-dirac}) can be rewritten in a more explicit form 
\begin{eqnarray}
(\hat{H}_0,\hat{\chi})_D&=&\frac{i}{\hbar}[\hat{H}_0,\hat{\chi}]-\frac{1}{2}
\left(\{\hat{H}_0,\hat{\chi}\}-\{\hat{\chi},\hat{H}_0\}\right)
\nonumber \\
&+&\frac{1}{2}\left(\sum_{\bar{\alpha},\bar{\beta}=1}^{2l}
\{\hat{H}_0,\xi_{\bar{\alpha}}\}({\bf C}^{-1})_{\bar{\alpha}\bar{\beta}}
\{\xi_{\bar{\beta}},\hat{\chi}\} \right.\nonumber\\
&-&\left.\sum_{\bar{\mu},\bar{\nu}=1}^{2l}\{\hat{\chi},\xi_{\bar{\mu}}\}
({\bf C}^{-1})_{\bar{\mu}\bar{\nu}}\{\xi_{\bar{\nu}},\hat{H}_0\}\right)
\;,\label{diracbraket-bis}
\end{eqnarray}
where indices with an overbar have been introduced to
indicate constraints.
The first two terms on the right hand side of Eq.~(\ref{diracbraket-bis})
give the standard quantum-classical bracket as given if Ref.~\cite{kakkacicco}.
The other two terms in the right hand side of Eq.~(\ref{diracbraket-bis})
pertain to the quantum-classical Dirac bracket and were not analyzed
in previous works.
In order to evaluate these terms, one has to remind the definition
of ${\bf C}^{-1}$ in Eq.~(\ref{C^-1})
and the fact that
\begin{eqnarray}
\frac{\partial\mbox{\boldmath$\xi$}}{\partial R}&=&
\left(
\frac{\partial\mbox{\boldmath$\sigma$}}{\partial R} , 
\frac{P}{M}\cdot\frac{\partial^2\mbox{\boldmath$\sigma$}}{\partial R\partial R}
 \right)
\\
\frac{\partial\mbox{\boldmath$\xi$}}{\partial P_i}&=&
\left({\bf 0},\frac{1}{M}\frac{\partial\mbox{\boldmath$\sigma$}}{\partial R}
\right)
\;.\label{holonomic}
\end{eqnarray}
Then consider
\begin{eqnarray}
&\sum_{\bar{\alpha},\bar{\beta}=1}^{2l}&
\{\hat{H}_0,\xi_{\bar{\alpha}}\}({\bf C}^{-1})_{\bar{\alpha}\bar{\beta}}
\{\xi_{\bar{\beta}},\hat{\chi}\} 
\nonumber\\
&&=\sum_{\bar{\alpha},\bar{\beta}=1}^{l}
\{\hat{H}_0,\sigma_{\bar{\alpha}}\}
({\bf Z}^{-1}\mbox{\boldmath$\bf\Gamma$}{\bf Z}^{-1})_{\bar{\alpha}\bar{\beta}}
\{\sigma_{\bar{\beta}},\hat{\chi}\}
\nonumber\\
&&
-\{\hat{H}_0,\sigma_{\bar{\alpha}}\}({\bf Z}^{-1})_{\bar{\alpha}\bar{\beta}}
\{\dot{\sigma}_{\bar{\beta}},\hat{\chi}\}
\nonumber\\
&&
+\{\hat{H}_0,\dot{\sigma}_{\bar{\alpha}}\}
({\bf Z}^{-1})_{\bar{\alpha}\bar{\beta}}
\{\sigma_{\bar{\beta}},\hat{\chi}\}
\label{HxiCxixhi_1}
\;.
\end{eqnarray}
Recalling that 
$\{\hat{H}_0,\sigma_{\bar{\alpha}}\}=-\dot{\sigma}_{\bar{\alpha}}=0$,
Eq.~(\ref{HxiCxixhi_1}) becomes
\begin{eqnarray}
&\sum_{\bar{\alpha},\bar{\beta}=1}^{2l}&
\{\hat{H}_0,\xi_{\bar{\alpha}}\}({\bf C}^{-1})_{\bar{\alpha}\bar{\beta}}
\{\xi_{\bar{\beta}},\hat{\chi}\} 
\nonumber\\
&&=\sum_{\bar{\alpha},\bar{\beta}=1}^l\{\hat{H}_0,\dot{\sigma}_{\bar{\alpha}}\}
({\bf Z}^{-1})_{\bar{\alpha}\bar{\beta}}
\{\dot{\sigma}_{\bar{\beta}},\hat{\chi}\}
\;.
\end{eqnarray}
Analogously, the last term in the right hand side
of Eq.~(\ref{diracbraket-bis}) is
\begin{eqnarray}
&\sum_{\bar{\mu},\bar{\nu}=1}^{2l}&
\{\hat{\chi},\xi_{\bar{\mu}}\}({\bf C}^{-1})_{\bar{\mu}\bar{\nu}}
\{\xi_{\bar{\nu}},\hat{H}_0\} \nonumber\\
&&=
-\sum_{\bar{\mu},\bar{\nu}=1}^l
\{\hat{\chi},\sigma_{\bar{\mu}}\}({\bf Z}^{-1})_{\bar{\mu}\bar{\nu}}
\{\dot{\sigma}_{\bar{\nu}},\hat{H}_0\}
\;.\label{chixiCxiHi_1}
\end{eqnarray}
Using Eqs.~(\ref{HxiCxixhi_1}) and~(\ref{chixiCxiHi_1}),
the Dirac quantum-classical bracket can be rewritten as
\begin{eqnarray}
(\hat{H}_0,\hat{\chi})_D&=&\frac{i}{\hbar}[\hat{H}_0,\hat{\chi}]-\frac{1}{2}
\left(\{\hat{H}_0,\hat{\chi}\}-\{\hat{\chi},\hat{H}_0\}\right)
\nonumber \\
&+&\frac{1}{2}\sum_{\bar{\mu},\bar{\nu}=1}^l
\left(
\{\hat{H}_0,\dot{\sigma}_{\bar{\mu}}\}({\bf Z}^{-1})_{\bar{\mu}\bar{\nu}}
\{\sigma_{\bar{\nu}},\hat{\chi}\}
\right.\nonumber\\
&+&\left.\{\hat{\chi},\sigma_{\bar{\nu}}\}({\bf Z}^{-1})_{\bar{\mu}\bar{\nu}}
\{\dot{\sigma}_{\bar{\nu}},\hat{H}_0\}
\right)
\;.\label{diracbraket-tris}
\end{eqnarray}
The last two terms in the right hand side of Eq.~(\ref{diracbraket-tris})
can be written more explicitly as
\begin{eqnarray}
&&\frac{1}{2}\sum_{\bar{\mu},\bar{\nu}=1}^l
\left(
\{\hat{H}_0,\dot{\sigma}_{\bar{\mu}}\}({\bf Z}^{-1})_{\bar{\mu}\bar{\nu}}
\{\sigma_{\bar{\nu}},\hat{\chi}\}\right.
\nonumber\\
&&+\left.\{\hat{\chi},\sigma_{\bar{\nu}}\}({\bf Z}^{-1})_{\bar{\mu}\bar{\nu}}
\{\dot{\sigma}_{\bar{\nu}},\hat{H}_0\}
\right)
\nonumber\\
&&=
\frac{1}{2}\sum_{\bar{\mu},\bar{\nu}=1}^l
\left[\left(
\frac{1}{M}
\frac{\partial\hat{H}_0}{\partial R}
\cdot\frac{\partial\sigma_{\bar{\mu}}}{\partial R}
-\frac{P}{M}\otimes\frac{P}{M}\cdot\frac{\partial^2\sigma_{\bar{\mu}}}{\partial R
\partial R}\right)
\right.\nonumber\\
&&\times
({\bf Z}^{-1})_{\bar{\mu}\bar{\nu}}\frac{\partial\sigma_{\bar{\nu}}}{\partial R}
\frac{\partial\hat{\chi}}{\partial R}
-\frac{\partial\hat{\chi}}{\partial P}\frac{\partial\sigma_{\bar{\nu}}}
{\partial R}({\bf Z}^{-1})_{\bar{\mu}\bar{\nu}}
\nonumber\\
&&\times\left.
\left(\frac{P}{M}\otimes\frac{P}{M}\cdot
\frac{\partial^2\sigma_{\bar{\nu}}}{\partial R\partial R}
- \frac{1}{M}
\frac{\partial\sigma_{\bar{\nu}}}{\partial R}
\frac{\partial\hat{H}_0}{\partial R}
\right)\right]\nonumber\\
\end{eqnarray}
With the above equations, the quantum-classical Dirac bracket can be written
finally as
\begin{eqnarray}
&&(\hat{H}_0,\hat{\chi})_D=\frac{i}{\hbar}[\hat{H}_0,\hat{\chi}]-\frac{1}{2}
\left(\{\hat{H}_0,\hat{\chi}\}-\{\hat{\chi},\hat{H}_0\}\right)
\nonumber \\
&&+
\frac{1}{2}\sum_{i,j}^{xyz}
\sum_{\bar{\mu},\bar{\nu}=1}^l
\left[\left(\frac{1}{M}
\frac{\partial\hat{H}_0}{\partial R}
\frac{\partial\sigma_{\bar{\mu}}}{\partial R}
-\frac{P}{M}\otimes\frac{P}{M}
\frac{\partial^2\sigma_{\bar{\mu}}}{\partial R\partial R}\right)
\right.\nonumber\\
&&\times
({\bf Z}^{-1})_{\bar{\mu}\bar{\nu}}\frac{\partial\sigma_{\bar{\nu}}}{\partial R}
\frac{\partial\hat{\chi}}{\partial R}
-\frac{\partial\hat{\chi}}{\partial P}\frac{\partial\sigma_{\bar{\nu}}}
{\partial R}({\bf Z}^{-1})_{\bar{\mu}\bar{\nu}}
\nonumber\\
&&\times\left.
\left(\frac{P}{M}\otimes\frac{P}{M}
\frac{\partial^2\sigma_{\bar{\nu}}}{\partial R\partial R}
- \frac{1}{M}
\frac{\partial\sigma_{\bar{\nu}}}{\partial R}
\frac{\partial\hat{H}_0}{\partial R}
\right)\right]\nonumber\\
\label{diracbraket-poker}
\end{eqnarray}
Equation~(\ref{diracbraket-poker}) is in a form suited
to be represented in the adiabatic basis.
In such a basis one can write
\begin{eqnarray}
\langle\alpha;R|(\hat{H}_0,\hat{\chi})_D|\alpha';R\rangle
&=&\sum_{\beta\beta'}i{\cal L}^D_{\alpha\alpha',\beta\beta'}\chi^{\beta\beta'}
\nonumber\\
&=&\sum_{\beta\beta'}\left(
i{\cal L}_{\alpha\alpha',\beta\beta'}
+i{\cal L}^{con}_{\alpha\alpha',\beta\beta'}
\right) \chi^{\beta\beta'}
\label{icalLD}
\;,\nonumber\\
\end{eqnarray}
where
\begin{eqnarray}
\sum_{\beta\beta'}
i{\cal L}_{\alpha\alpha',\beta\beta'} \chi^{\beta\beta'}
&=&
\langle\alpha;R|
\left[\frac{i}{\hbar}[\hat{H}_0,\hat{\chi}]
\right.\nonumber\\
&-&\left.\frac{1}{2}
\left(\{\hat{H}_0,\hat{\chi}\}-\{\hat{\chi},\hat{H}_0\}\right)
\right] |\alpha';R\rangle
\;.
\nonumber\\
\end{eqnarray}
The Liouville operator $i{\cal L}_{\alpha\alpha',\beta\beta'}$
was given in Ref.~\cite{kakkacicco}
\begin{eqnarray}
i{\cal L}_{\alpha\alpha',\beta\beta'}
&=&i\omega_{\alpha\alpha'}\delta_{\alpha\beta}\delta_{\alpha'\beta'}
+iL_{\alpha\alpha'}\delta_{\alpha\beta}\delta_{\alpha'\beta'}
-J_{\alpha\alpha',\beta\beta'}
\;,
\nonumber\\
\end{eqnarray}
where
$\omega_{\alpha\alpha'}=\hbar^{-1}(E_{\alpha}(R)-E_{\alpha'}(R))$,
\begin{eqnarray}
iL_{\alpha\alpha'}&=&
\frac{P}{M}\frac{\partial}{\partial R}
+\frac{1}{2}\left(F^{\alpha}+F^{\alpha'}\right)
\frac{\partial}{\partial P}
\end{eqnarray}
is a classical-like Liouville operator
which makes quantum-classical variables evolve
on a constant energy surface with Helmann-Feynman
forces given by $(1/2)(F^{\alpha}+F^{\alpha'})$, and
\begin{eqnarray}
J_{\alpha\alpha',\beta\beta'}
&=&-\delta_{\alpha'\beta'}d_{\alpha\beta}
\left[\frac{P}{M}+\frac{\hbar}{2}
\omega_{\alpha\beta}\frac{\partial}{\partial P}\right]
\nonumber\\
&-&\delta_{\alpha\beta}d_{\alpha'\beta'}^*
\left[\frac{P}{M}+\frac{\hbar}{2}
\omega_{\alpha'\beta'}\frac{\partial}{\partial P}\right]
\label{J}
\end{eqnarray}
is an off-diagonal operator, realizing nonadiabatic transitions,
which is defined in terms of the nondiabatic coupling vector 
$d_{\alpha\beta}=\langle\alpha|\partial/\partial R|\beta\rangle$.
The operator $i{\cal L}^{con}_{\alpha\alpha',\beta\beta'}$
which imposes the constraints in the quantum-classical dynamics is
\begin{eqnarray}
&\sum_{\beta\beta'}&
i{\cal L}^{con}_{\alpha\alpha',\beta\beta'} \chi^{\beta\beta'}
\nonumber\\
&&=\frac{1}{2}\sum_{\bar{\mu}\bar{\nu}}
\langle\alpha;R|\left(\frac{1}{M}
\frac{\partial\hat{H}_0}{\partial R}
\frac{\partial\sigma_{\bar{\mu}}}{\partial R}
-\frac{P}{M}\otimes\frac{P}{M}\cdot
\frac{\partial^2\sigma_{\bar{\mu}}}{\partial R\partial R}\right)
\nonumber\\
&&\times
({\bf Z}^{-1})_{\bar{\mu}\bar{\nu}}\frac{\partial\sigma_{\bar{\nu}}}{\partial R}
\frac{\partial\hat{\chi}}{\partial R}|\alpha';R\rangle
\nonumber\\
&&-\frac{1}{2}\sum_{\bar{\mu}\bar{\nu}}\langle\alpha;R|
\frac{\partial\hat{\chi}}{\partial P}\frac{\partial\sigma_{\bar{\nu}}}
{\partial R}({\bf Z}^{-1})_{\bar{\mu}\bar{\nu}}
\nonumber\\
&&\times
\left(\frac{P}{M}\otimes\frac{P}{M}\cdot
\frac{\partial^2\sigma_{\bar{\nu}}}{\partial R\partial R}
-\frac{1}{M}
\frac{\partial\sigma_{\bar{\nu}}}{\partial R}
\cdot
\frac{\partial\hat{H}_0}{\partial R}
\right)|\alpha';R\rangle\;.\nonumber\\
\end{eqnarray}
By defining 
$F^{\alpha\beta}=-\langle\alpha;R|\partial\hat{H}_0/\partial R|\beta;R\rangle$
and $F^{\alpha}=-\partial E_{\alpha}(R)\partial R$, using
$F^{\alpha\beta}=F^{\alpha}\delta_{\alpha\beta}+\hbar\omega_{\alpha\beta}
d_{\alpha\beta}$, adding and subtracting the term
\begin{eqnarray}
\frac{1}{2}(F^{\alpha}+F^{\alpha'})\sum_{\bar{\mu},\bar{\nu}=1}^l
\frac{1}{M}
\frac{\partial\sigma_{\bar{\mu}}}{\partial R}
({\bf Z}^{-1})_{\bar{\mu}\bar{\nu}}
\frac{\partial\sigma_{\bar{\nu}}}{\partial R}
\frac{\partial }{\partial P}\delta_{\alpha\beta}\delta_{\alpha'\beta'}
\;,
\end{eqnarray}
and doing some algebra one finally obtains
\begin{eqnarray}
i{\cal L}^{con}_{\alpha\alpha',\beta\beta'}
&=&-\sum_{\bar{\nu}=1}^l\lambda_{\bar{\nu}}^{\alpha\alpha'}
\frac{\partial\sigma_{\bar{\nu}}}{\partial R}
\frac{\partial}{\partial P}\delta_{\alpha\beta}\delta_{\alpha'\beta'}
\nonumber\\
&&-\frac{1}{2M}\sum_{\bar{\mu},\bar{\nu}=1}^l
\left(\hbar\omega_{\alpha\beta}d_{\alpha\beta}
\delta_{\alpha'\beta'}\right.\nonumber\\
&+&\left.\hbar\omega_{\alpha'\beta'}d_{\alpha'\beta'}^*\delta_{\alpha\beta}
\right) \cdot
\frac{\partial\sigma_{\bar{\mu}}}{\partial R}
({\bf Z}^{-1})_{\bar{\mu}\bar{\nu}}
\frac{\partial\sigma_{\bar{\beta}}}{\partial R}\cdot\frac{\partial}{\partial P}
\;,\label{iLcon}
\nonumber\\
\end{eqnarray}
where the $\lambda_{\bar{\nu}}^{\alpha\alpha'}$, which
are the quantum-classical Lagrange multipliers on the energy
surface $(1/2)(E_{\alpha}+E_{\alpha'})$, are defined as
\begin{eqnarray}
\lambda_{\bar{\nu}}^{\alpha\alpha'}
&=&
\sum_{\bar{\mu},\bar{\nu}=1}^l\left(\frac{P}{M}\otimes\frac{P}{M}\cdot
\frac{\partial^2\sigma_{\bar{\mu}}}{\partial R\partial R}
\right.\nonumber\\
&+&\left.\frac{1}{2M}
(F^{\alpha}+F^{\alpha'})
\frac{\partial\sigma_{\bar{\mu}}}{\partial R}
\right)({\bf Z}^{-1})_{\bar{\mu}\bar{\nu}}
\;.
\end{eqnarray}
The first term on the right hand side of Eq.~(\ref{iLcon}) defines
a diagonal operator 
\begin{eqnarray}
iL^{con}_{\alpha\alpha'}&=&
-\sum_{\bar{\nu}=1}^l\lambda_{\bar{\nu}}^{\alpha\alpha'}
\frac{\partial\sigma_{\bar{\nu}}}{\partial R}
\frac{\partial}{\partial P}\;,
\end{eqnarray}
whose action is to enforce the constraints while time evolution
takes place adiabatically on the energy surface 
$(1/2)(E_{\alpha}+E_{\alpha'})$. The other two terms in the right hand
side of Eq.~(\ref{iLcon}) defines an off-diagonal operator
\begin{eqnarray}
J_{\alpha\alpha',\beta\beta'}^{con}
&=&
\frac{1}{2M}\sum_{\bar{\mu},\bar{\nu}=1}^l
\left(\hbar\omega_{\alpha\beta}d_{\alpha\beta}
\delta_{\alpha'\beta'}\right.\nonumber\\
&+&\left.\hbar\omega_{\alpha'\beta'}d_{\alpha'\beta'}^*\delta_{\alpha\beta}
\right)
\cdot
\frac{\partial\sigma_{\bar{\mu}}}{\partial R}
({\bf Z}^{-1})_{\bar{\mu}\bar{\nu}}
\frac{\partial\sigma_{\bar{\beta}}}{\partial R}\cdot\frac{\partial}{\partial P}
\nonumber\\
\end{eqnarray}
which couples the constrained dynamics to quantum transitions
between the adiabatic states. Its effect is, on the one hand, to modify
the probability that quantum transitions take place and, on the other one,
to realize the back-reaction of the quantum transitions on the
constrained momenta.
It is convenient to finally cast
the quantum-classical Dirac-Liouville operator of Eq.~(\ref{icalLD})
in the following form
\begin{eqnarray}
i{\cal L}^D_{\alpha\alpha',\beta\beta'}&=&
\left(i\omega_{\alpha\alpha'}
+iL_{\alpha\alpha'}^D\right)\delta_{\alpha\beta}\delta_{\alpha'\beta'}
-J^D_{\alpha\alpha',\beta\beta'}\;,
\nonumber\\
\label{dirac-liouville}
\end{eqnarray}
where
\begin{eqnarray}
iL_{\alpha\alpha'}^D&=&iL_{\alpha\alpha'}+iL_{\alpha\alpha'}^{con}
\nonumber\\
&=&
\frac{P}{M}\frac{\partial}{\partial R}
+\frac{1}{2}\left(F^{\alpha}+F^{\alpha'}\right)
\frac{\partial}{\partial P}
\nonumber\\
&-&\sum_{\bar{\nu}=1}^l\lambda_{\bar{\nu}}^{\alpha\alpha'}
\frac{\partial\sigma_{\bar{\nu}}}{\partial R}
\frac{\partial}{\partial P}\;,
\end{eqnarray}
and
\begin{eqnarray}
J^D_{\alpha\alpha',\beta\beta'}&=&
J_{\alpha\alpha',\beta\beta'}+J_{\alpha\alpha',\beta\beta'}^{con}
\nonumber\\
&=&-\left(\frac{P}{M}\cdot d_{\alpha\beta}\right)
\left[1+\frac{\hbar}{2}
\frac{{\omega}^D_{\alpha\beta}d_{\alpha\beta}}
{\left(\frac{P}{M}\cdot d_{\alpha\beta}\right)}
\cdot\frac{\partial}{\partial P}
\right]
\delta_{\alpha'\beta'}
\nonumber\\
&-&\left(\frac{P}{M}\cdot d^*_{\alpha'\beta'}\right)
\left[1+\frac{\hbar}{2}
\frac{{\omega}^D_{\alpha'\beta'}d^*_{\alpha'\beta'}}
{\left(\frac{P}{M}\cdot d^*_{\alpha'\beta'}\right)}
\cdot\frac{\partial}{\partial P}\right]
\delta_{\alpha\beta}
\;,\label{J^D}
\nonumber\\
\end{eqnarray}
where the constrained frequency is
\begin{eqnarray}
\omega^D_{\alpha\beta}&=&
\omega_{\alpha\beta}
\left(1+\sum_{\bar{\mu},\bar{\nu}=1}^l
\frac{1}{M}\frac{\partial\sigma_{\bar{\mu}}}{\partial R}
({\bf Z}^{-1})_{\bar{\mu}\bar{\nu}}
\frac{\partial\sigma_{\bar{\nu}}}{\partial R}\right)
\;. \label{omega^D}
\end{eqnarray}
The time evolution of any dynamical variable is given
obviously by
\begin{eqnarray}
\frac{\partial}{\partial t}\chi^{\alpha\alpha'}
&=&\sum_{\beta\beta'}i{\cal L}^D_{\alpha\alpha',\beta\beta'}\chi^{\beta\beta'}
\;.
\label{timeevol}
\end{eqnarray}

\subsection{MOMENTUM-JUMP APPROXIMATION FOR CONSTRAINED SYSTEMS}

When performing numerical calculations on systems with
many degrees of freedom~\cite{sergi,sergi2,sergi3,donal}
the action of the operator in Eq.~(\ref{J})
is usually evaluated within the momentum-jump approximation~\cite{simu}.
This approximation can be derived easily for the
operator in Eq.~(\ref{J^D}) defined in terms of the constrained frequency
$\omega^D_{\alpha\beta}$ in Eq.~(\ref{omega^D}).
In analogy with what shown in Ref.~\cite{simu}, in order to derive the
momentum-jump approximation for constrained system,
one can  first consider one of the two terms in the right hand side
of Eq.~(\ref{J^D})
\begin{eqnarray}
\frac{\hbar}{2}\omega_{\alpha\beta}^D d_{\alpha\beta}
\left(\frac{P}{M}\cdot d_{\alpha\beta}\right)
\frac{\partial}{\partial P}
&=&\hbar\omega_{\alpha\beta}^DM\frac{\partial}
{\partial(P\cdot\hat{d}_{\alpha\beta})^2}
\;,\nonumber\\
\end{eqnarray}
where $\hat{d}_{\alpha\beta}$ denotes the unit vector along the direction
of $d_{\alpha\beta}$.
Then perform the approximation
\begin{eqnarray}
1+\hbar\omega_{\alpha\beta}^DM\frac{\partial}
{\partial(P\cdot\hat{d}_{\alpha\beta})^2}
&\approx&
\exp\left[\hbar\omega_{\alpha\beta}^DM\frac{\partial}
{\partial(P\cdot\hat{d}_{\alpha\beta})^2}\right]\;,
\nonumber\\
\end{eqnarray}
and write the operator $J^D_{\alpha\alpha',\beta\beta'}$
as the operator $J^{D,M-J}_{\alpha\alpha',\beta\beta'}$
evaluated in the momentum-jump approximation
\begin{eqnarray}
J^{D,M-J}_{\alpha\alpha',\beta\beta'}
&=&
-\left(\frac{P}{M}\cdot d_{\alpha\beta}\right)
\exp\left[\hbar M \omega^D_{\alpha\beta}
\frac{\partial}{\partial(P\cdot\hat{d}_{\alpha\beta})^2}\right]
\delta_{\alpha'\beta'}
\nonumber\\
&-&\left(\frac{P}{M}\cdot d^*_{\alpha'\beta'}\right)
\exp\left[\hbar M \omega^D_{\alpha'\beta'}
\frac{\partial}{\partial(P\cdot\hat{d}^*_{\alpha'\beta'})^2}\right]
\delta_{\alpha\beta}
\;.\label{J^DM-J}
\nonumber\\
\end{eqnarray}
The action of any of the operators on the right hand side 
of Eq.~(\ref{J^DM-J}) on an arbitrary function of momenta
is a translation (or jump) of the momenta itself.
For example,
\begin{eqnarray}
\exp\left[\hbar\omega_{\alpha\beta}^DM\frac{\partial}
{\partial(P\cdot\hat{d}_{\alpha\beta})^2}\right]
f(P)&=&f(P+\Delta P)\;, \nonumber\\
\end{eqnarray}
with
\begin{eqnarray}
\Delta P&=&
sign(P\cdot\hat{d}_{\alpha\beta})
\sqrt{(P\cdot\hat{d}_{\alpha\beta})^2+\hbar M\omega^D_{\alpha\beta}}
-(P\cdot\hat{d}_{\alpha\beta})\;.
\nonumber\\
\label{DeltaP}
\end{eqnarray}
Equation~(\ref{DeltaP}) illustrates the effect of the momentum-jump operator
on a function of constrained momenta. It differs from that given in Ref.~\cite{simu}
and used, for example, in Refs.~\cite{sergi,sergi2,sergi3}
because of the constrained frequency $\omega_{\alpha\beta}^D$
defined in Eq.~(\ref{omega^D}). This result, which should have been used
in the nonadiabatic calculations of Ref.~\cite{hannakakka},
is reasonable because constrained momenta cannot be scaled
as unconstrained momenta. As a matter of fact, if one follows the empirical
procedure of applying
the unconstrained momentum-jump of Ref.~\cite{simu}, as it was done
in Ref.~\cite{hannakakka}, and uses the RATTLE procedure~\cite{rattle}
to impose the constraints on momenta then a result different from that
given in Eq.~(\ref{DeltaP}) is obtained.
Instead, Equation~(\ref{DeltaP}), with the momentum-jump operator for the
constrained system in Eq.~(\ref{J^DM-J}), was derived correctly from
the fundamental Dirac quantum-classical formalism and it provides
the correct formula for constrained systems.

\section{Stationary Dirac Density Matrix}\label{sec:stationary}

Write the quantum-classical Dirac
density matrix of a constrained system as $\hat{\rho}_D(X)$.
The average of any operator $\hat{\chi}$
can be calculated from
\begin{equation}
\langle \hat{\chi}\rangle
={\rm Tr}'\int dX~\hat{\rho}_D\hat{\chi}(t)
={\rm Tr}'\int dX~\hat{\rho}_D
\exp\left(i{\cal L}^D t\right)\hat{\chi} \;.
\end{equation}
The action of $\exp\left(i{\cal L}^D t\right)$ can be transferred
from $\hat{\chi}$ to $\hat{\rho}_D$ by using the cyclic invariance 
of the trace and integrating by parts the term coming from
the  classical Dirac brackets.
One can write
\begin{equation}
i{\cal L}^D=\frac{i}{\hbar}\left[\hat{H}_0,\dots \right]-\frac{1}{2}
\left(\{\hat{H}_0,\dots\}_D-\{\dots,\hat{H}_0\}_D\right).
\end{equation}
In this equation the classical Dirac bracket terms
are written
\begin{eqnarray}
\{\hat{H}_0,\dots \}_D-\{\dots,\hat{H}_0\}_D
&=&
\sum_{i,j=1}^{2N}\left(
\frac{\partial\hat{H}_0}{\partial X_i}
{\cal B}_{i j}^D
\frac{\partial \ldots}{\partial X_j}
\right.
\nonumber \\
&-&\left.
\frac{\partial \dots}{\partial X_i}
{\cal B}_{i j}^D
\frac{\partial \hat{H}_0}{\partial X_j}\;.
\right)
\end{eqnarray}
When integrating by parts the right hand side,
one obtains terms proportional to the compressibility
\begin{eqnarray}
\hat{\kappa}_0^D
&=&\sum_{i,j=1}^{2N}\frac{\partial {\cal B}_{i j}^D}{\partial X_i}
\frac{\partial \hat{H}_0}{\partial X_j}
\;.
\end{eqnarray}
Using Eqs.~(\ref{B^D}) and~(\ref{C^-1}), in analogy
with Ref.~\cite{bdispettosa}, one finds easily that
\begin{eqnarray}
\kappa_0^D&=&-\frac{d}{dt}\ln det{\bf Z}\;.
\label{k_0^Dconstr}
\end{eqnarray}
Because of the compressibility in Eq.~(\ref{k_0^Dconstr}),
it turns out that
the Dirac quantum-classical Liouville operator
is not hermitian
\begin{equation}
    \left(i\hat{\cal L}^D\right)^{\dag} 
    =-i\hat{\cal L}^D-\kappa_0^D\;.
\end{equation}
The average value can then be written as
\begin{equation}
\langle \hat{\chi}\rangle
={\rm Tr}'\int dX~\hat{\chi} \exp\left[-(i{\cal L}^D+
\kappa_0^D) t\right]\hat{\rho}_D
\;.
\end{equation}
The quantum-classical Dirac density matrix
evolves under the equation
\begin{eqnarray}
\frac{\partial}{\partial t}\hat{\rho}_D&=&-\frac{i}{\hbar}
\left[\hat{H}_0,\hat{\rho}_D\right]+\frac{1}{2}
\left(\{\hat{H}_0,\hat{\rho}_D\}_D-\{\hat{\rho}_D,\hat{H}_0\}_D\right)
\nonumber \\
&-&\kappa_0^D\hat{\rho}_D\;.
\label{eq:qc-dme}
\end{eqnarray}
The stationary density matrix $\hat{\rho}_{De}$ is defined by
\begin{equation}
i{\cal L}^D\hat{\rho}_{De}+\kappa_0^D\hat{\rho}_{De}=0\;.
\label{eq:qm-case}
\end{equation}
To find the explicit expression of $\hat{\rho}_{De}$
one can follow Ref.~\cite{nciccokakka},
expand the density matrix in powers of $\hbar$
\begin{equation}
\hat{\rho}_{De}=\sum_{n=0}^{\infty}\hbar^n\hat{\rho}_{De}^{(n)}
\;,
\end{equation}
and look for an explicit solution in the adiabatic basis.
In such a basis the Dirac-Liouville operator is expressed by
Eq.~(\ref{dirac-liouville}) and the Hamiltonian is given by
\begin{eqnarray}
H_0^{\alpha}&=&\frac{P^2}{2M}+ E_{\alpha}(R) \;.
\end{eqnarray}
Thus, one obtains an infinite set of equations corresponding to
the various powers of $\hbar$ 
\begin{eqnarray}
iE_{\alpha\alpha'}\rho_{De}^{(0)\alpha\alpha'}&=&0
\;,
\label{stat-n=0}
\\
iE_{\alpha\alpha'}\rho_{De}^{(n+1)\alpha\alpha'}&=&
-iL_{\alpha\alpha'}^D \rho_{De}^{(n)\alpha\alpha'}
- \kappa_0^D
\rho_{De}^{(n)\alpha\alpha'}
\nonumber \\
&+&\sum_{\beta\beta'}J^D_{\alpha\alpha',\beta\beta'}
\rho_{De}^{(n)\beta\beta'}
~ (n\ge 1)\;.
\label{stat-n+1}
\end{eqnarray}
As shown in Ref.~\cite{nciccokakka},
in order to ensure that a solution can be found
by recursion, one must discuss the solution of Eq.~(\ref{stat-n+1})
when calculating the diagonal elements $\rho_{De}^{(n)\alpha\alpha}$ in terms
of the off-diagonal ones $\rho_{De}^{(n)\alpha\alpha'}$.
To this end, using
$\rho_{De}^{\prime(n)\alpha\alpha'}=(\rho_{De}^{\prime(n)\alpha'\alpha})^*$,
$J_{\alpha\alpha,\beta\beta'}^D=J_{\alpha\alpha,\beta'\beta}^{D*}$
and the fact that $J^D_{\alpha\alpha,\beta\beta}=0$ when
a real basis is chosen,
it is useful to re-write Eq.~(\ref{stat-n+1}) in the form
\begin{equation}
(iL_{\alpha\alpha}^D+\kappa_0^D)
\rho_{De}^{(n)\alpha\alpha}=\sum_{\beta>\beta'}2{\cal R}
\left(J_{\alpha\alpha,\beta\beta'}^D
\rho_{De}^{(n)\beta\beta'}\right)\;.
\label{stat-condition}
\end{equation}
One has~\cite{bdispettosa} 
$(-iL_{\alpha\alpha}^D-\kappa_0^D)^{\dag}
=iL_{\alpha\alpha}^D$.
The right hand side of this equation
can expressed by means
of the classical Dirac bracket in Eq.~(\ref{dirac-bracket}).
It follows that
$H_0^{\alpha}$ and any general function $f(H_0^{\alpha})$
are constants of motion under the action of $iL_{\alpha\alpha}^D$.
Because of the presence of a non-zero phase space compressibility,
integrals over phase space must be taken using the invariant 
measure~\cite{bdispettosa,tuckerman}
\begin{equation}
d{\cal M}=\exp(-w_D)dRdP\;,
\end{equation}
where $w_D=\int dt\kappa_0^D=det{\bf Z}$
is the indefinite integral of the compressibility.
To insure that a solution to Eq.~(\ref{stat-condition})
exists one must invoke the theorem
of Fredholm alternative, requiring that the 
right-hand side of Eq.~(\ref{stat-condition}) 
be orthogonal to the null space of 
$(iL_{\alpha\alpha}^D)^{\dagger}$~\cite{hilbert}.
The null-space of this operator
consists of functions of the form~\cite{b2}
$f(H_0^{\alpha})$,
where $f(H_0^{\alpha})$ can be any function
of the adiabatic Hamiltonian $H_0^{\alpha}$.
Thus the condition to be satisfied is
\begin{equation}
\int d{\cal M}\sum_{\beta>\beta'}2{\cal R}
\left(J^D_{\alpha\alpha,\beta\beta'}\rho_{De}^{(n)\beta\beta'}\right)
f(H_0^{\alpha})=0 \;.
\label{fredholm}
\end{equation}
To this end,
there is no major difference with the proof given in Ref.~\cite{nciccokakka}:
$2{\cal R}\left(J^D_{\alpha\alpha,\beta\beta'}\rho_{De}^{(n)\beta\beta'}\right)$
and $f(H_0^{\alpha})$ are respectively 
an odd and an even function of $P$; 
this guarantees the validity of Eq.~(\ref{fredholm}).
Thus, one can write the formal solution of Eq.~(\ref{stat-condition}) as
\begin{equation}
\rho_{De}^{(n)\alpha\alpha}
=(iL_{\alpha\alpha}^D+\kappa_0^D)^{-1}
\sum_{\beta>\beta'}2{\cal R}
\left(J_{\alpha\alpha,\beta\beta'}^D\rho_{De}^{(n)\beta\beta'}\right)\;,
\label{eq:sol1}
\end{equation}
and the formal solution of Eq.~(\ref{stat-n+1}) for 
$\alpha\neq\alpha'$ as
\begin{eqnarray}
\rho_{De}^{(n+1)\alpha\alpha'}&=&
\frac{i}{E_{\alpha\alpha'}}
(iL_{\alpha\alpha'}^D+\kappa_0^D)
\rho_{De}^{(n)\alpha\alpha'}
\nonumber \\
&&
-\frac{i}{E_{\alpha\alpha'}}
\sum_{\beta\beta'}J^D_{\alpha\alpha',\beta\beta'}\rho_{De}^{(n)\beta\beta'}
\;.
\label{eq:sol2}
\end{eqnarray}
Equations~(\ref{eq:sol1}) and~(\ref{eq:sol2}) allows one to calculate
$\rho_{De}^{\alpha\alpha'}$ to all orders in $\hbar$
once $\rho_{De}^{(0)\alpha\alpha'}$ is given.
This order zero term is obtained by the solution of
$(iL_{\alpha\alpha}^D+\kappa_0^D)
\rho_{De}^{(0)\alpha\alpha}=0$. All higher order terms
are obtained by the action of $E_{\alpha\alpha'}$, the imaginary unit 
$i$ and $J^D_{\alpha\alpha'\beta\beta'}$ (involving factors of 
$d_{\alpha\alpha'}$, $\omega^D_{\alpha\alpha'}$, $P$,
and derivatives with respect to $P$.

One can find a stationary solution
to order $\hbar$
by considering the first two equations of the set
given by Eqs.~(\ref{stat-n=0}) and (\ref{stat-n+1}):
\begin{eqnarray}
\left[\hat{H}_0,\hat{\rho}_{De}^{(0)}\right]& =&0  
\qquad \qquad  (n=0)\;,
\label{n=0}
\\
i\left[\hat{H}_0,\hat{\rho}_{De}^{(1)}\right]&=&
+\frac{1}{2}\left(
\{\hat{H}_0,\hat{\rho}_{De}^{(0)}\}_D
-
\{\hat{\rho}_{{\rm N}e}^{(0)},\hat{H}_0\}_D\right)
\nonumber\\
&-&\frac{1}{2}[\hat{\kappa}_0^D,\hat{\rho}_{De}^{(0)}]_+
\qquad  (n=1)\;.
\label{n=1}
\end{eqnarray}

For the ${\cal O}(\hbar^0)$ term one can make the ansatz
\begin{equation}
\hat{\rho}_{De}^{(0)\alpha\beta}=\frac{1}{Q}det{\bf Z}
\delta\left({\cal C}-H^{\alpha}_0\right)
\delta(\mbox{\boldmath$\xi$})\delta_{\alpha\beta}
\;, \label{ansatz}
\end{equation}
where $Q$ is
\begin{eqnarray}
Q&=&\sum_{\alpha}\int d{\cal M}~\delta(\mbox{\boldmath$\xi$})
\delta\left({\cal C}-H^{\alpha}_0\right)
\end{eqnarray}
and $\delta(\mbox{\boldmath$\xi$})$ is a compact notation for
\begin{eqnarray}
\delta(\mbox{\boldmath$\xi$})&=&
\prod_{\bar{\nu}=1}^l\delta(\sigma_{\bar{\nu}})
\prod_{\bar{\mu}=1}^l\delta(\dot{\sigma}_{\bar{\mu}})\;.
\end{eqnarray}
The following expression for the order $\hbar$ term is obtained
\begin{eqnarray}
\hat{\rho}_{De}^{(1)\alpha \beta}
&=&-i
\frac{P}{M}d_{\alpha \beta}\hat{\rho}_{De}^{(0) \beta}
\left[\frac{1-e^{-\beta(E_{\alpha}-E_{\beta})}}{E_{\beta}-E_{\alpha}}
+\frac{\beta}{2}
\right.
\nonumber \\
&& \left.
\left(1+e^{-\beta(E_{\alpha}-E_{\beta})}\right)
 \right]
 \label{ansatz2}
\end{eqnarray}
for the ${\cal O}(\hbar)$ term.

Equations~(\ref{ansatz}) and (\ref{ansatz2}) give the explicit form
of the stationary solution of the Dirac Liouville equation
up to order ${\cal O}(\hbar)$. This stationary solution
\underline{must} be used when calculating equilibrium
averages in quantum-classical systems with holonomic
constraints on the classical variables.

\section{Linear Response Theory}\label{sec:lrt}

Consider a perturbed quantum-classical Hamiltonian
\begin{eqnarray}
\hat{H}(t)=\hat{H}_0+\hat{H}_I(t)= \hat{H}_0-\hat{A}{\cal F}(t)\;.
\end{eqnarray}
Define the perturbed Liouville operator 
\begin{eqnarray}
i{\cal L}^D_I(t)&=& \frac{i}{\hbar}
\left[\hat{H}_I(t),\ldots\right]\nonumber\\
&-&\frac{1}{2}
\left(\{\hat{H}_I(t),\ldots\}_D-\{\ldots,\hat{H}_I(t)\}_D\right)
\;.
\end{eqnarray}
In general the perturbation can bring an additional term
to the compressibility of phase space
\begin{eqnarray}
\hat{\kappa}_I^D&=&-\frac{\partial{\cal B}_{ij}^D}{\partial X_i}
\frac{\partial\hat{A}}{\partial X_j}{\cal F}(t)
=-\hat{\kappa}_A^D{\cal F}(t)\;.
\end{eqnarray}
Using Eqs.~(\ref{B^D}) and~(\ref{C^-1}), one finds
\begin{eqnarray}
\kappa_A^D&=&-\left(\frac{\partial\ln det{\bf Z}}{\partial R}
+\sum_{\bar{\nu},\bar{\mu}=1}^lZ_{\bar{\nu}\bar{\mu}}
\frac{\partial Z_{\bar{\nu}\bar{\mu}}^{-1}}{\partial R}\right)
\cdot\frac{\partial \hat{A}}{\partial P}\;.
\label{k_A^D}
\end{eqnarray}
In this case, the pertubed Liouville operator is not hermitian
\begin{eqnarray}
(i{\cal L}^D_I(t))^{\dag}&=&-i{\cal L}^D_I(t)- 
\frac{1}{2}[\hat{\kappa}_I^D(t),\ldots]_+ \;,
\end{eqnarray}
where $[\ldots,\ldots]_+$ denotes the anticommutator
\begin{eqnarray}
[\hat{\kappa}_I^D,\hat{\rho}_{D}]_+
&=&
\left[\begin{array}{cc} \hat{\kappa}_I^D & \hat{\rho}_{D}\end{array}\right]
\cdot\left[\begin{array}{cc} {\bf 0} & {\bf 1}\\ {\bf 1} & {\bf 0}
\end{array}\right]\cdot
\left[\begin{array}{c} \hat{\kappa}_I^D \\ \hat{\rho}_{D}\end{array}\right]
\nonumber\\
&=&\hat{\kappa}_I^D\hat{\rho}_{D}++\hat{\rho}_{D}\hat{\kappa}_I^D\;.
\end{eqnarray}
The constrained evolution of the density matrix is obtained
from Eq.~(\ref{eq:qc-dme}) by replacing $\hat{H}_0$ by
$\hat{H}(t)$
\begin{eqnarray}
\frac{\partial}{\partial t}\hat{\rho}_D(t)&=&-\frac{i}{\hbar}
\left[\hat{H}(t),\hat{\rho}_D(t)\right]\nonumber\\
&+&\frac{1}{2}
\left(\{\hat{H}(t),\hat{\rho}_D(t)\}_D-\{\hat{\rho}_D(t),\hat{H}_0\}_D\right)
\nonumber \\
&-&\kappa_0^D\hat{\rho}_D(t)
-\frac{1}{2}[\hat{\kappa}_I^D(t),\hat{\rho}_D(t)]_+\;.
\label{eq:qc-dme-pert}
\end{eqnarray}

Assuming that the perturbed density matrix
is $\hat{\rho}_D(t)=\hat{\rho}_{De}+\Delta\rho_D(t)$
and that the system was in equilibrium in the distant past,
linear response theory gives
\begin{eqnarray}
\Delta\rho_D(t)&=&
-\int_{-\infty}^td\tau\exp[-i{\cal L}^{D\dag}(t-\tau)]
i{\cal L}_I^{D\dag}(\tau)\hat{\rho}_{De}
\;.  \nonumber\\
\label{lr_1}
\end{eqnarray}
If one defines
\begin{eqnarray}
i{\cal L}^D_A(t)&=&
\frac{i}{\hbar}
\left[\hat{A}(t),\ldots\right]\nonumber\\
&-&\frac{1}{2}
\left(\{\hat{A}(t),\ldots\}_D-\{\ldots,\hat{A}\}_D\right)
\;,
\end{eqnarray}
then $i{\cal L}^D_I(t)=-{\cal F}(t)i{\cal L}^D_A$
and Eq.~(\ref{lr_1}) becomes
\begin{eqnarray}
\Delta\rho_D(t)&=&
\int_{-\infty}^td\tau{\cal F}(\tau)\exp[-i{\cal L}^{D\dag}(t-\tau)]
i{\cal L}_A^{D\dag}\hat{\rho}_{De}\nonumber\\
&=&-
\int_{-\infty}^td\tau{\cal F}(\tau) e^{-i{\cal L}^{D\dag}(t-\tau)}
\left(i{\cal L}_A^D\hat{\rho}_{De}\right.\nonumber\\
&+&\left.\frac{1}{2}
[\hat{\kappa}_A^D,\hat{\rho}_{De}]_+\right)
\;.
\label{lr_2}
\end{eqnarray}
The non-equilibrium average of any quantum-classical operator $\hat{B}(X)$
can be calculated over the density matrix $\hat{\rho}_D(t)$,
$\overline{B}=Tr'\int dX \hat{B}(X)\hat{\rho}_D(t)$
in order to determine the response of the system
to the external force.
\begin{eqnarray}
\overline{\Delta B}(t)&=&Tr'\int dX \hat{B}(X)\Delta\rho_D(t)
\nonumber\\
&=&\int_{-\infty}^t d\tau\Phi_{BA}(t-\tau){\cal F}(\tau)
\;,
\end{eqnarray}
where the response function is defined as
\begin{eqnarray}
\Phi_{BA}(t)&=&-Tr'\int dX \hat{B}(X;t)
\left(i{\cal L}_A^D\hat{\rho}_{De}
+\frac{1}{2}
 [\hat{\kappa}_A^D,\hat{\rho}_{De}]_+\right)
\;.\nonumber\\\label{phiba}
\end{eqnarray}
Equation~(\ref{phiba}) gives the response function for quantum-classical
systems with holonomic constraints on the classical variables.
In analogy with the purely classical case,
analized in Ref.~\cite{bdispettosa}, 
if $\hat{A}(X)$ depends from the momenta $P$
the response function contains two contributions
in addition to the expression
of $\Phi_{BA}(t)$ given in Ref.~\cite{nciccokakka}.
One of this contributions arises
evidently from $\hat{\kappa}_A^D$ and the other comes
from the factor $det{\bf Z}$ contained in the expression
for $\hat{\rho}_{De}$.

Because a perturbation operator which depends only
on particle positions is usually adopted 
in order to derive expressions for quantum-classical rate constants
(typically the Heaviside function),
the rate formulas derived in Refs~\cite{sergi,sergi2,hannakakka}
also applies to a constrained system
with the exception that the correct constrained stationary matrix,
as derived in Sec~\ref{sec:stationary}, must be used.
However, different calculations and different perturbation
operators may require the evaluation of the additional
terms of the response function here derived
and one must be aware of their existence.

\section{CONCLUSIONS}\label{sec:concl}

In this paper the theory of quantum-classical systems has been generalized
in order to treat rigorously situations where the classical degrees of freedom
must obey holonomic constraints. The formalism here presented has been obtained
by unifying the classical Dirac bracket with the quantum-classical bracket
in matrix form. In this way, a Dirac quantum-classical formalism, which conserves
the constraints exactly, has been
introduced and then used consistently to formulate the dynamics and the statistical
mechanics of quantum-classical systems with holonomic constraints.
A first result has been the derivation of the correct momentum-jump
approximation which takes into account that, when a quantum transition occurs,
the momenta cannot cannot be scaled as if they were unconstrained
because, instead, the holonomic constraints must be satisfied.
Moreover, the Dirac quantum-classical bracket allows one to derive easily 
linear response theory. This has shown that the rigorous response function
of constrained systems contains non-trivial terms, which were already noted by
this author in classical mechanics, arising from the action of the perturbation
operator on the phase space measure of unperturbed constrained systems
and from the compression of phase space which may be caused by the perturbation
itself. These terms are zero if the external perturbation is coupled only
to the position coordinates of the classical degrees of freedom.

If one considers with a wider perspective this work and that of Ref.~\cite{b3},
dealing with the introduction of non-Hamiltonian commutators in quantum mechanics,
it can be realized that a unified formalism for defining generalized dynamics
in quantum-classical systems is now available. 
There are reasonable expectations of employing in the future
specific forms of such generalized dynamics in order to attack the problem
of long time numerical integration of quantum-classical dynamics.

\begin{flushleft}
{\bf ACKNOWLEDGMENT}
\end{flushleft}
The author is grateful to Professor Raymond Kapral for having introduced him
to the theory of quantum-classical systems.


\end{document}